\documentclass[a4,aps,prl,twocolumn,groupeaddress,nofootinbib]{revtex4}
\usepackage{amsmath}

\usepackage{graphicx}
\usepackage{latexsym}
\usepackage{array}
\usepackage{xcolor}
\usepackage[normalem]{ulem}

\newcolumntype{P}[1]{>{\centering\arraybackslash}p{#1}} 
\newcolumntype{M}[1]{>{\centering\arraybackslash}m{#1}}
\input{epsf}

\newcommand{\correc}{\textcolor{black}}
\newcommand{\syl}{\textcolor{black}}
\newcommand{\thie}{\textcolor{black}}

\newcommand{\tim}{\textcolor{black}}
\newcommand{\timn}{\textcolor{black}}

\newcommand{\gern}{\textcolor{black}}

\begin{document}
\title{\correc{Succession} of resonances to achieve internal wave turbulence}
\author{G\'eraldine Davis$^1$, Timoth\'ee Jamin$^1$, Julie Deleuze$^1$, Sylvain Joubaud$^{1,2}$, Thierry Dauxois$^1$}
\affiliation{1. Univ Lyon, ENS de Lyon, Univ Claude Bernard, CNRS, Laboratoire de Physique, Lyon, France\\
2. Institut Universitaire de France (IUF)}
\date{\today}

\begin{abstract}
We study experimentally the interaction of nonlinear internal waves in a stratified fluid confined in a trapezoidal tank. The set-up has been \correc{designed to produce internal wave turbulence from monochromatic and polychromatic forcing through three processes.} The first is a linear \correc{transfer}  in wavelength obtained by \correc{wave} reflection on inclined slopes, leading to an internal wave attractor \correc{which has a broad wavenumber spectrum}. 
\correc{Second is the broad banded time-frequency spectrum of the trapezoidal geometry,}  \correc{as shown by the impulse response of the system.}
\correc{The third one is} a nonlinear \correc{transfer}  in frequencies and wavevectors via triadic  interactions, which 
results at large forcing amplitudes in a power law decay of the \correc{wavenumber power} spectrum.
This first experimental spectrum of internal wave turbulence displays a $k^{-3}$ behavior.
\end{abstract}

\maketitle

\correc{Internal gravity waves propagate within density stratified fluids} \correc{moving under the influence of buoyancy forces~\cite{SutherlandBook}.}
 \correc{Recently, they have been actively studied in particular because of their importance to mixing and transport in the ocean. For example, a coordinated observational campaign} has been performed
in the South China Sea~\cite{alfordpeacocknature}, \correc{which is} well-known to \correc{contain breaking internal waves} with amplitudes up to 200~m.
Their generation through the interaction between tides and bathymetry~\cite{LuzonExp}, their propagation and 
 instability~\cite{DauxoisARFM2018}, 
\correc{and their} interaction  with oceanic currents  are  \correc{just some of the outstanding dynamics being observed}. \correc{In particular, the study of mixing by breaking internal waves is relevant for the understanding \correc{of}  biological processes such as the vertical redistribution of zooplankton from the deep ocean}~\cite{plankton} \correc{and the regeneration of} the coral reef of Dongsha \correc{atoll}~\cite{dongsharef}.

In the nonlinear regime, \correc{stratified} fluid systems may develop \correc{turbulence} simultaneously \correc{due to} wave\correc{s}  and 
vortices~\cite{Campagne2015}. 
Describing the coexistence of
\correc{each process} is a challenge in itself. 
If stratified  turbulence has been \correc{actively} studied  (see~\cite{Colm} and references therein), 
wave turbulence \correc{for internal waves} is \correc{a relatively unexplored phenomenon}. 
Wave turbulence describes physical systems with a large number of dispersive and nonlinear interacting waves~\cite{NazarenkoBook}
and has been applied to gravity~\cite{DeikeBerhanuFalcon}, capillary~\cite{Herbertetal} and inertial waves~\cite{SharonNaturePhysics},
\correc{as well as} waves in magnetised fluids~\cite{magnetized} and  in
 elastic plates~\cite{Miqueletal}.
New applications \correc{have} recently emerged in condensed matter (superfluid Helium and Bose-Einstein condensates), in
nonlinear optics~\cite{picozzi} and, most recently, in the study of gravitational waves in the early universe~\cite{GaltierNazarenko}.
Internal  
waves are \correc{distinct from} these waves~\cite{Tabak}, \correc{owing} to
their unusual dispersion relation. In this letter,  
 we present an experimental set-up that allows us to \gern{observe} efficient nonlinear energy transfers \correc{in} frequenc\correc{y} and  \correc{wavenumber} and \correc{so determine the signature of}
  internal wave turbulence.

\begin{figure}[b!]
\includegraphics[width=\columnwidth]{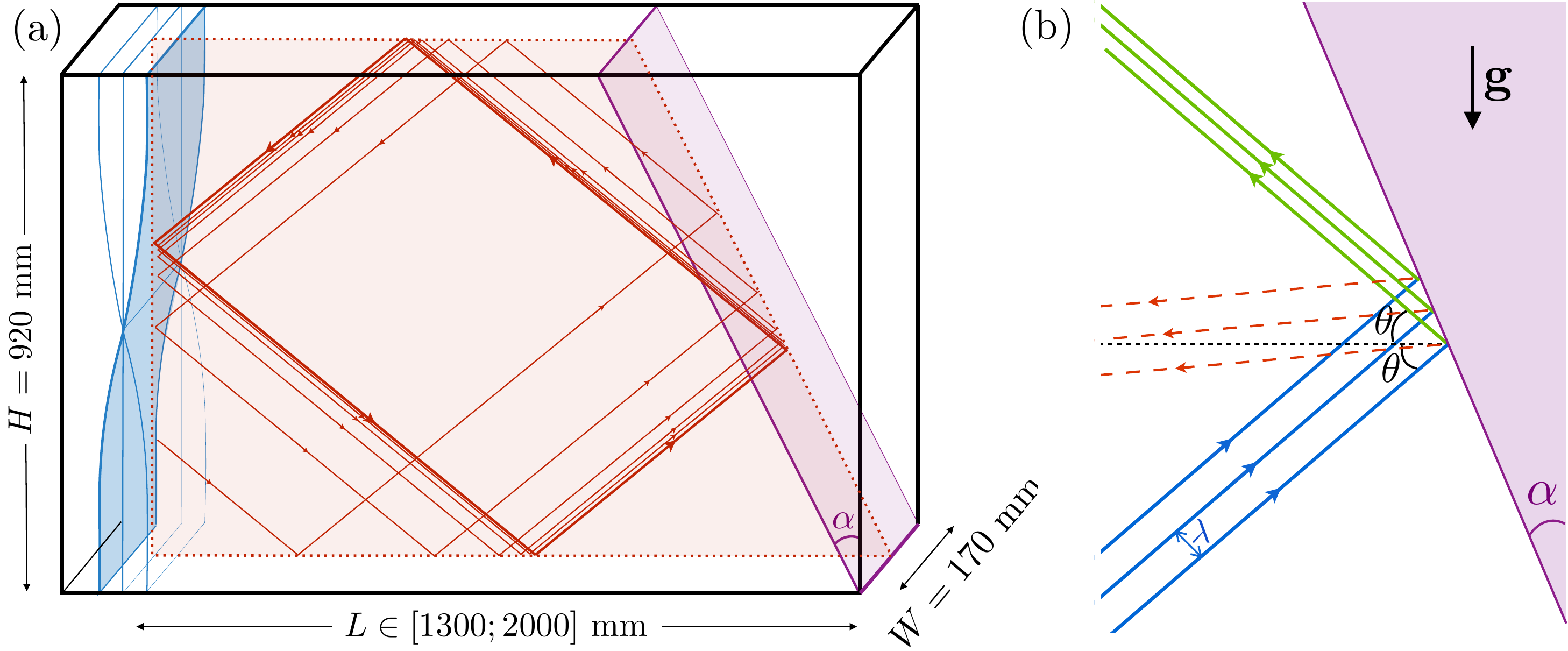}
\caption{a) Experimental set-up 
with the generator \correc{on the left}  of the tank and a slope inclined with an angle~$\alpha$ \correc{on the right}.
An example of ray tracing from a single point located on the left hand wall and corresponding to an oscillation at a given frequency~$\omega=\pm N\sin\theta$ 
is shown in the case of a linearly stratified fluid with constant $N$-value. b) Schematic reflection on a slope of an incident 
internal gravity beam (in blue) with two wavelengths \syl{$\lambda$} in the transversal direction.
The reflected beam does not follow the Snell-Descartes prediction (in dashed red) but 
keeps the same angle with respect to the horizontal, or with the gravity $\mathbf{g}$, as shown by the green beam.
}
\label{fig:fig1_syl}
\end{figure}

The experimental set-up, sketched in Fig.~\ref{fig:fig1_syl}(a), 
is a confined trapezoidal domain filled with a linear stratified fluid~of density $\rho(z)$, \correc{in which $z$ is the vertical coordinate.}
\correc{Introducing} the gravity $g$ and a reference density $\rho_{\textrm{ref}}$, \correc{the strength of the stratification is characterized by} 
the buoyancy frequency $N=\sqrt{-({g}/{\rho_{\textrm{ref}}}){\partial \rho}/{\partial z}}$.
 \correc{In all the experiments, this} is of the order $1$~rad/s. The energy is injected at  large scale by means 
 of a vertical  boundary oscillating \correc{horizontally} around its mid horizontal axis, 
 \correc{with} a half cosine shape $a(t)\cos\left(\pi z/H\right)$\correc{, in which $a(t)$ is the maximum horizontal 
 displacement and $H$ is the depth.} \correc{The volume is thus kept constant.} \correc{The classical experiments are \correc{performed with} a quasi-monochromatic forcing, 
 $a(t) = a_0\sin(\omega t)$. Here, we also examine impulsive \correc{and polychromatic} forcings.}  The resulting velocity field\correc{s are} measured using the classical particle image velocimetry \gern{(PIV)} method.

\correc{The dispersion relation of internal waves, $\omega= N\sin\left\vert\theta\right\vert$,} \correc{is such that the frequency~\correc{$\omega$} sets 
the angle~\correc{$\theta$}  of propagation of internal wave beams and also
the ratio of the horizontal to vertical wavenumber.} 
Reflections of internal waves on vertical or horizontal walls are analogous 
to \correc{optical} reflection, preserving
 the angle with respect to the normal of the boundary.
On the contrary, reflection on a slope interestingly
leads   to a ray with the same angle with the horizontal \correc{as the incident ray} (in absolute value);
this is a simple consequence of the preservation of the frequency $\omega$ after reflection
with  the dispersion relation.  Figure~\ref{fig:fig1_syl}(b) shows that this simple mechanism is even more 
interesting when considering a beam since  both its width and its wavelength are reduced after reflection (notice that, for internal waves, the wavevector is orthogonal to the ray). As 
the group velocity is proportional to the wavelength, \correc{energy} conservation leads to  efficient energy \correc{focusing of the beam}.

\begin{figure}[h!]
\includegraphics[width=\columnwidth]{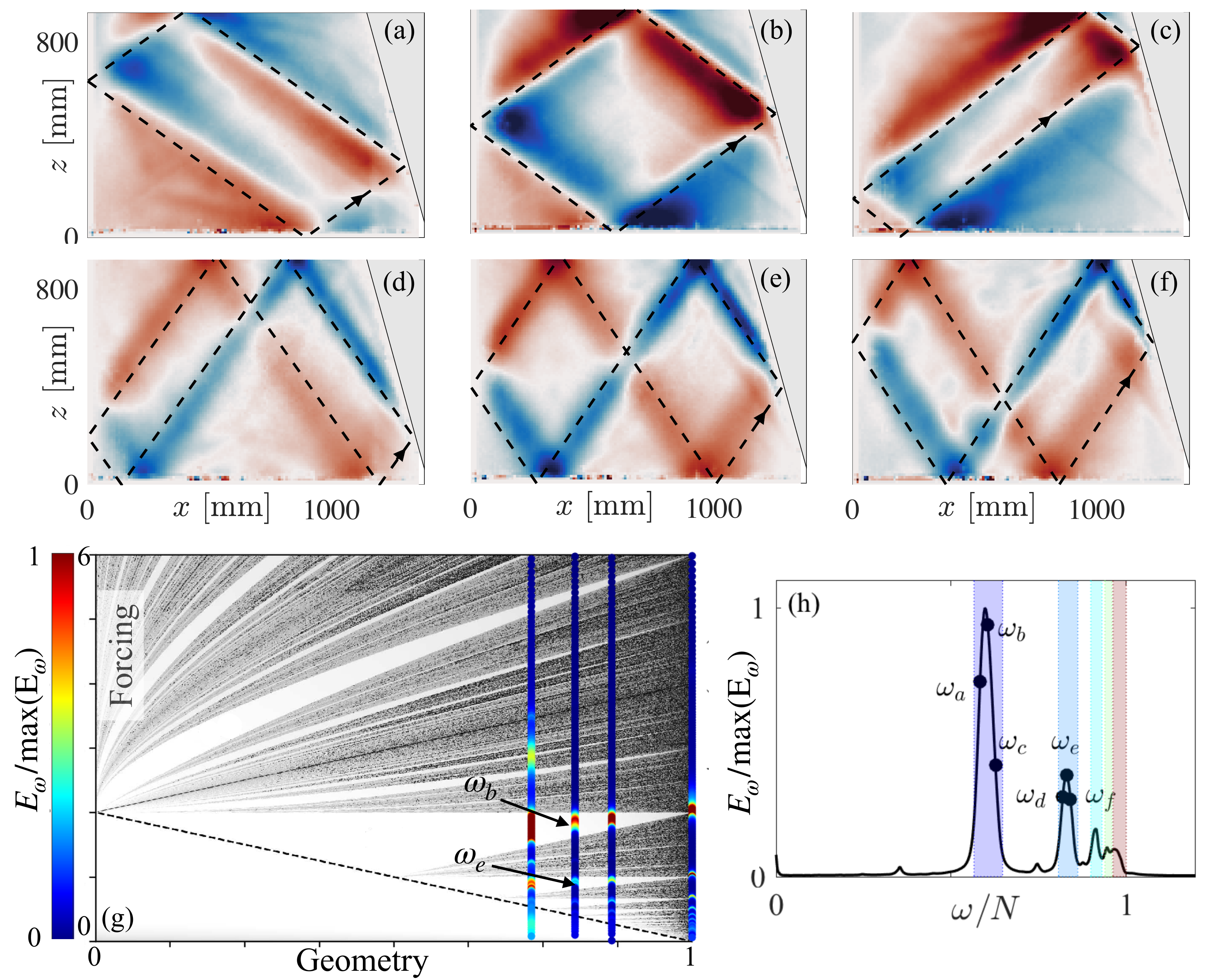}
\caption{\correc{a-f) Spatial patterns obtained after an initial impulsive kick when the experimental signal is filtered around different frequencies shown with symbols in (h).
One recovers (1,1) attractors  ({a to c}) and (2,1) attractors ({d to f}). The theoretically expected attractors for the geometry and corresponding frequencies \gern{are} depicted with the dashed line. g) \correc{In grey scale,} Lyapunov exponents of the trapezoidal domain in the inviscid limit (from~\cite{MaasNature}) as a function 
of the geometry ({$1-(2H/L)\tan \alpha$}) and of the forcing ({$(2H/L)\sqrt{(N/\omega)^2-1}$}) \correc{which is decreasing while increasing the frequency} . White regions correspond to strongly convergent attractors. h)  Energy spectrum measured experimentally via PIV 
after an initial kick of the generator for $\alpha=15.5^\circ$. Vertical colored bands emphasize the theoretical frequency bands of internal wave attractor{s} in the stationary regime. The same spectra for~$\alpha=0$, 11.1, 15.5 and 21.4$^\circ$ are superimposed
on (f) as colored vertical lines, from right to left (corresponding graphs are presented in  the Supplemental Material~\cite{SupplementalMaterial})}}
\label{fig:Lyapunov}
\end{figure}

\correc{After multiple reflections,}
internal waves generated at a given frequency
 concentrate on a closed loop~\cite{MaasLam1995}\correc{, as illustrated by a single ray traced in Fig.~\ref{fig:fig1_syl}(a).
 The shape and rotational direction of the so-called internal wave attractor
(IWA) are independent of the initial emitting point and thus of the spatial structure of the forcing.}
\correc{The IWA can be seen as a limit cycle, a prominent word in nonlinear physics, which arises from linear theory.} 
\correc{D}ifferent angles of propagation  \correc{set by different forcing frequencies} 
lead to different attractors with simple or more complicated shapes: \correc{the dashed lines} in Figure~\ref{fig:Lyapunov}\correc{(a-f)} presents 
a few $(m,n)$ \correc{theoretical} attractors, with $m$ reflection{s} on the top and $n$ reflections on the slope.

\correc{Due to} viscosity,
\correc{IWAs have finite} width \correc{as a consequence of} the balance 
between 
 \correc{focusing} and dissipation \correc{as shown in Fig.~\ref{fig:Lyapunov}(a-f)}~\cite{Beckebanze2018,Davis2019}. 
The energy \correc{concentration}, \correc{which is} large just after  focusing \correc{upon} reflection \correc{from the slope}, is progressively dissipated
along the \correc{length} of the attractor before being focused again by the
slope. Recent experiments of IWA  in two \correc{nontrivial} three-dimensional set-ups~\cite{Grimaud2018}
have additionally shown that this is more than a beautiful mathematical curiosity.

It is important to \correc{emphasize} that \correc{this energy}  \correc{focusing}
\correc{with corresponding} 
transfer to small scales arises from linear theory.
\correc{T}his is \correc{the} first important ingredient in the context of wave turbulence 
as it increases the number of waves with different \correc{wavenumbers}
in interaction.

Owing to energy focusing, the reflected beam has a larger amplitude and is therefore  more prone  to reach the threshold for 
 triadic resonant instability (TRI)~\cite{DauxoisARFM2018,SED2013}: 
a beam with primary frequency $\omega$ and wavevector ${\mathbf k}$ \correc{can excite from background noise} two subharmonic waves with frequencies~$\omega_\pm$ 
and wavevectors~${\mathbf k}_\pm$, satisfying the temporal and spatial resonance 
conditions $\omega=\omega_++\omega_-$   and ${\mathbf k}={\mathbf k}_++{\mathbf k}_-$. 
 \correc{\correc{Any} subharmonic} may \correc{also} become unstable through TRI and/or interact with \correc{an}other \correc{pre-existing} wave to generate a third one,
the latter process being without any amplitude threshold. 
One thus obtains a necessary ingredient for wave turbulence: a physical mechanism providing 
\correc{multiple nonlinear interactions} between waves of various wavelengths and frequencies.

{\em The kicked attractor experiment}.
 Before building on \correc{this method} to achieve internal wave turbulence, 
\correc{we consider the special case of transient impulsive forcing by analogy with} musicians \correc{striking} a tuning fork 
that resonates at a specific constant pitch. 

If \correc{the internal wave} generator 
is set into motion with \correc{time-dependent amplitude} $a(t)$, which  is  non-zero \correc{i}n a short interval $\delta t\ll 2\pi /N$\correc{, then a broad} frequenc\correc{y} \correc{spectrum} will be simultaneously excited\correc{.} \correc{We then consider the impulse response of the \correc{trapezoidal} domain.}

\correc{Besides,} \correc{plotting} the Lyapunov exponent quantifies the exponential divergence of rays issued from close initial points.
Figure~\ref{fig:Lyapunov}\correc{(g)} presents the theoretical prediction~\cite{MaasNature}
in the inviscid limit. White tongues correspond to domains 
of existence of the different $(m,n)$ attractors. 

The impulse response of the different experiments are analyzed 
by considering the \correc{frequency} content of the kinetic energy 
\gern{$E_\omega(\omega)= \int \left ( \vert\tilde{u}(x,z,\omega)\vert^2+\vert\tilde{w}(x,z,\omega)\vert^2 \right) \mathrm{d}x \mathrm{d}z$},
 computed from the Fourier transform of the two components of the velocity field and \gern{integrated} over $(x,z)$ plane.
\correc{Such energy spectra probe complete vertical lines in Fig.~\ref{fig:Lyapunov}(g).}
In \correc{the} absence of any inclined slope ($\alpha=0$)\correc{, when the geometry is 1 in Fig.~\ref{fig:Lyapunov}(g)}, one gets a set of well\correc{-}peaked discrete modes as shown by the rightmost vertical line
(see the \gern{energy} spectrum
 in the Supplemental Material{~\cite{SupplementalMaterial}}).
This is expected because attractor\correc{s, which require a sloping boundary, cannot}
exist in \correc{a} rectangular domain\correc{.} 
\correc{O}ne thus recovers  the discrete set of resonance frequencies of the rectangle. The width of these peaks observed experimentally is due to viscous effects.
On the contrary, for \correc{non-}\correc{zero}  $\alpha$\correc{,} Fig.~\ref{fig:Lyapunov}\correc{(g)} shows that
discrete modes are replaced by  vertical bands located within the domain 
of existence of the different IWA\correc{s}. 
{These bands of \correc{peak} energy are \correc{plotted} in Fig.~\ref{fig:Lyapunov}\correc{(h)} for a given geometry.}
Filtering the velocity field at given frequenc{ies} allows us to disentangle the different responses to the initial \correc{impulsive} kick.
 Figures~\ref{fig:Lyapunov}\correc{(a-f)} present different attractors that appear when the signal is filtered around frequencies belonging to the
(1,1) and (2,1) tongues of Fig.~\ref{fig:Lyapunov}\correc{(a)}. Note that only IWA\correc{s} with short perimeters
 ({\em i.e.} with small $m$ and $n$ values) \correc{are visible}
 in presence of damping.  

Introducing a slope therefore modifies the usual picture 
of a wave operator with discrete eigenmodes to some continuous spectrum, and understanding the linear response of these  systems is a non-trivial question~\cite{YvesEtLaure}.
By taking advantage of \correc{such frequency band within one of the tongue
we can efficiently force  the system in a polychromatic way,}
thus increasing the number of waves in interaction.
{As almost all regions of the physical domain will be covered by an IWA, this characteristic also allows us to inject energy rather homogeneously\correc{. Both properties are beneficial in the study of wave turbulence.}

{\em Beyond the linear regime.} 
\correc{We first consider} the transition to nonlinear dynamics \correc{resulting from} monochromatic forcing \correc{$a(t) =a_0\sin(\omega_0 t)$}. 
Inspired by the transition to wave turbulence observed numerically in harmonically 
forced elastic plates~\cite{Touzeetal}, the amplitude of the generator~$a_0$ is  gradually (from 2 to 10 mm) and slowly (over $750$ oscillating periods)
increased,
as shown in  Fig.}~\ref{fig:AnalogyTouze}(a).
The frequency has been chosen within the $(1,1)$ tongue.
The velocity field, whose horizontal component is presented in Fig.~\ref{fig:AnalogyTouze}(b-d) \correc{at three different times}, is analyzed with the
time-frequency \correc{function}, 
$S_{\mathbf{u}}(t,\omega) = 
\langle
\vert \int_{-\infty}^{+\infty} \textrm{d}t'\, h(t-t') e^{i\omega t'} u \vert^2 + 
\vert \int_{-\infty}^{+\infty} \textrm{d}t'\, h(t-t') e^{i\omega t'} w \vert^2 \rangle_{x,z}$, \correc{which is plotted} in Fig.~\ref{fig:AnalogyTouze}(e). 
\correc{The function $h$ is a hanning window \timn{of $80\,T_0$}. This time window is wide enough to resolve subharmonic frequencies and narrow enough to consider forcing amplitude constant within its duration (increase  of 0.8\timn{5}\,mm).}
The horizontal dark-red line at $\omega/N=0.62$ in Fig.~\ref{fig:AnalogyTouze}(e) corresponds to the forcing frequency.  One distinguishes 
several different regimes:  the {\em linear regime} (Fig.~\ref{fig:AnalogyTouze}(b)) with a monochromatic response leading to a \correc{well-defined single attractor}.  
 Figure~\ref{fig:AnalogyTouze}(c) corresponds  to a TRI-perturbed IWA\correc{:} {the two subharmonic waves \correc{(around $\omega_-/N=0.25$ and $\omega_+/N=0.37$)} \correc{apparent at this time}
in Fig.~\ref{fig:AnalogyTouze}(e) clearly satisfy the temporal resonance condition \gern{with $\omega_0$}.
\correc{We also observe $\omega_0+\omega_-$ and $\omega_0+\omega_+$, the signatures of the interaction of these subharmonics with the attractor.}
The third and turbulent-like regime (Fig.~\ref{fig:AnalogyTouze}(d)) is characterized by a broadband Fourier spectrum and a spatial pattern in which \correc{no single IWA is evident.}
When the amplitude is further increased \correc{the enrichment of the spectrum is progressive,}
contrary to numerical simulation\correc{s of the elastic plates}~\cite{Touzeetal}. 
\correc{Although} some peaks are still visible at large forcing (especially at the forcing frequency), 
a  continuous spectrum is \correc{nonetheless} established.
 \correc{\correc{We} estimate \correc{(see Supplemental Material~\cite{SupplementalMaterial}) that} the nonlinear interaction between waves \tim{typically} occurs at a time scale \tim{\timn{one to ten times} larger than}
  the linear one}\gern{, thus leading to a strong wave turbulent regime.}
  
\begin{figure}[t!]
\includegraphics[width=\columnwidth]{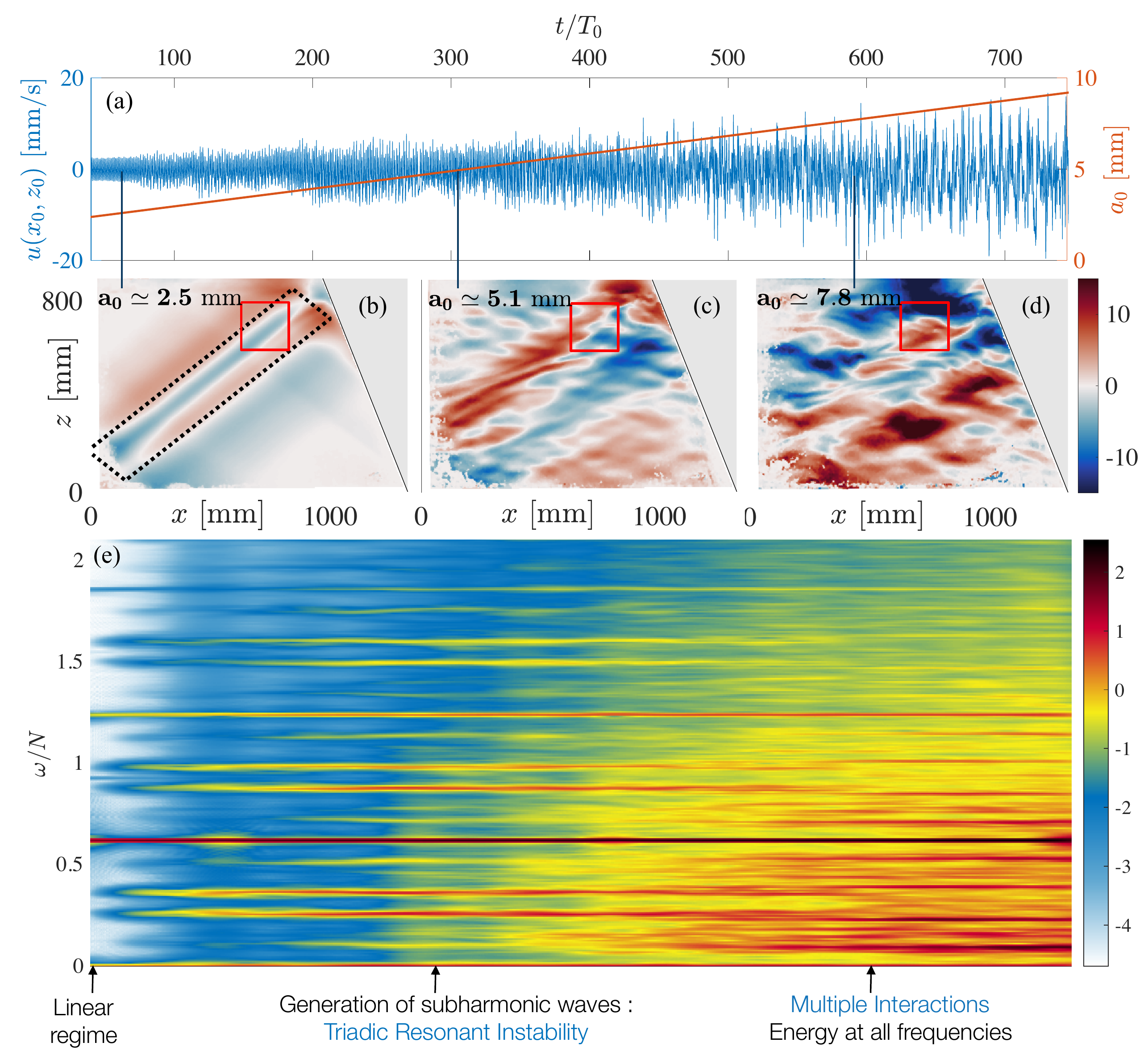}
\caption{a) Amplitude ramp (in red) and horizontal velocity component $u(x,z,t)$ measured in a point at the center of the red square
shown in (b). b-d) Three snapshots of the horizontal velocity fields \correc{(in mm/s)} measured when $a=2.5$, 5.1 and  7.8~mm.
e)~Time-frequency diagram  $ S_{\mathbf{u}}(t,\omega)$  of the velocity field of the red square domain (in logarithmic scale). $\alpha=21.4^\circ$, $L=1328$~mm, $N=0.7$ rad/s  and $\omega_0/N=0.62$.}
\label{fig:AnalogyTouze}
\end{figure}	

{\em Spatial Spectra}.\
Figure~\ref{fig:SpatialSpectrum}(a) presents the evolution of the \correc{wavenumber} \correc{power} spectrum $E_k (k)= {k}\left( \langle\vert\hat{u}(k,\theta,t)\vert^2+\vert\hat{w}(k,\theta,t)\vert^2\rangle_{\theta,t} \right)\timn{/\cal{A}}$ 
as the forcing amplitude is gradually increased. \correc{Here, $k=\vert {\mathbf k} \vert$ is the wavenumber\timn{, $\cal{A}$ is the area of the trapezoidal domain} and} $\hat{u}$ and $\hat{w}$ are the  Fourier transform\correc{s} of the components of the velocity field.}
 To improve the signal to noise ratio, the spectrum has been averaged over all angle\correc{s} $\theta$ and over many forcing periods, as indicated 
  by~$\langle \cdot \rangle_{\theta,t}$.
 The analysis has been performed
 only on the low frequency band ({\em i.e.} $\omega<N$),
 \correc{where propagative waves are predominant (see~\cite{BrouzetEPL} and Fig.~\ref{fig:Disp_relation} in the Supplemental Material~\cite{SupplementalMaterial}).}

 \begin{figure}[t!]
\includegraphics[width=0.98\columnwidth]{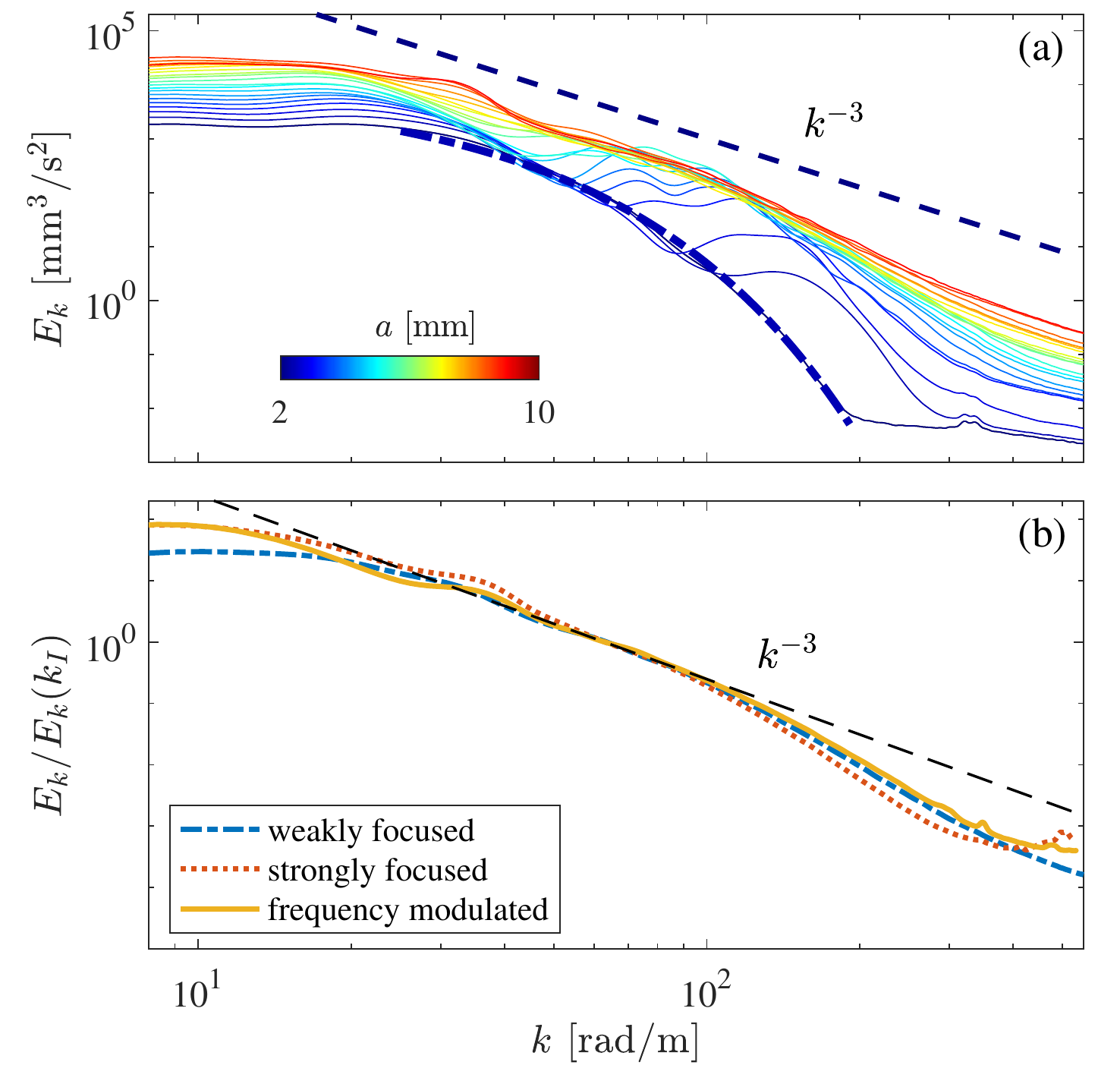}
\caption{(a) \correc{Power spectral} density 
\correc{ $E_k (k)= {k}\left( \langle\vert\hat{u}(k,\theta,t)\vert^2+\vert\hat{w}(k,\theta,t)\vert^2\rangle_{\theta,t} \right)\timn{/\cal{A}}$}
as a function of the wavenumber~$k$ for the experiment shown in Fig.~\ref{fig:AnalogyTouze}.
The spatial Fourier transform of the components of the velocity field $\hat{u}$ and $\hat{w}$ are  computed using a hanning window.
The different curves correspond to an increase of the amplitude~$a$ from 2 to 10 mm \gern{(that lasts for 750 periods), with an average over 35 forcing periods for each curve.} 
The thick dash-dotted blue line corresponds to a fit, whose exponential shape is predicted for a linear IWA.
(b) Stationary spectrum \timn{averaged over $100$ periods} for different forcings. \correc{The different experimental parameters are given in the Supplemental Material~\cite{SupplementalMaterial}.} In both panels, the dashed line shows the power law $E_k\simeq k^{-3}$. $k_I\simeq60$ rad/m is taken in the center of the inertial range.
}
\label{fig:SpatialSpectrum}
\end{figure}	

In the linear regime, IWA\correc{s} have been shown to have an exponential power spectrum \correc{$E_0\exp(-\beta k)$},
when dissipation due to the lateral walls dominates that of the bulk~\cite{Beckebanze2018}.
This behaviour is indeed observed at 
small amplitude ($a=2$ mm), as shown by the dashed blue line.
When $a$ is slightly increased,  the attractor is destabilized by TRI and a small \correc{peak} is visible around $k=150${~rad/m}\correc{, corresponding to a 4~cm wavelength.}
\correc{Such} subharmonic waves  have a wavelength smaller than the typical width of the attractor~\cite{DauxoisARFM2018}.
Further increasing~$a$,  the \correc{peak} not only moves towards smaller values of $k$ but also widens.

This \gern{first} observation can be explained as the coupling between  \gern{the} \correc{spectral} energy transfer \correc{due to the linear focusing} and the \correc{one} due to nonlinear interactions.
\correc{T}he  energy \correc{focusing} is  \correc{now} balanced \correc{both} by  dissipation \correc{and by extraction} via  TRI leading to a wider \correc{beam within the IWA}.
The primary wave\correc{,} having a larger wavelength,  generates secondary waves with larger wavelengths as well  (see Supplemental Material~\cite{SupplementalMaterial}).
Moreover, \correc{because} their group velocity $c_g=N\cos\theta/k$ \correc{is} larger, subharmonic waves \correc{rapidly} fill the whole tank.

At large forcing, these different ingredients lead to a richer spectrum that is compatible with a power law decay  $E_k\sim k^{-3}$, as shown in Fig.~\ref{fig:SpatialSpectrum}(a).
 \correc{T}he same exponent has been obtained in \correc{various numerical simulations with different forcing mechanisms}~\cite{PascaleJoelChantal,LeReun}.
  
\correc{We have gone on to examine these analyses for experiments with} different forcing and  geometrical parameters. 
We \correc{have} studied
a strongly focused  IWA that corresponds to a larger Lyapunov exponent than for the weakly focused IWA shown in Fig.~\ref{fig:SpatialSpectrum}(a)\correc{.}
\correc{T}aking advantage of the band structure 
revealed in the kicked attractor experiment, 
we \correc{also} forced 
the stratified fluid via a \correc{white noise filtered in the frequency range of}
 the (1,1) attractor shown in Fig.~\ref{fig:Lyapunov}{(h)}. 
In \correc{both} cases, the exponent -3 is robust as shown in Fig.~\ref{fig:SpatialSpectrum}(b). 

{\em Conclusion.} \correc{We have performed the first experimental measurements of internal wave turbulence.} The trapezoidal\correc{,} stably\correc{-}stratified domain \correc{is a robust} experimental set-up to study nonlinear\correc{ly} interacting internal waves.
Different forcings, \correc{whether} monochromatic or \correc{frequency modulated}, lead to
a \correc{power spectrum with a} well\correc{-}defined power law\correc{.} 

\correc{Future work will focus on examination of the frequency spectra.}
For frequenc\correc{ies} below the buoyancy frequency~$N$, one expects a \correc{$\omega^{-2}$ spectrum similar to} the 
Garrett and Munk spectrum \correc{of observed oceanic internal waves~{\cite{GarrettandMunk}}.}  \correc{This} is often used
as a representative statistical description of the internal
wave field in studies of nonlinear interaction,  despite 
\correc{only an approximate description}. \correc{For frequencies} above~$N$, a steeper $\omega^{-4}$ \correc{spectrum} has been recently reported in numerical simulations~\cite{Mininni}.
Whether or not both \correc{spectra are manifest} in experiments remains an open question.
\correc{T}o be closer to  oceanic \correc{circumstances}, \correc{experiments are being designed with} a less constrained geometrical set-up that would allow \correc{for} three-dimensional dynamics\correc{.}

{\em Acknowledgements.}
We are indebted to  C. Brouzet, E. Ermanyuk, C. Herbert, L. Maas\correc{,}  L. Saint Raymond \correc{and B. Sutherland} for useful discussions. 
This work was supported by the grant ANR-17-CE30-0003 (DisET) and by the LABEX iMUST (ANR-10-LABX-0064) of Universit\'e de Lyon, 
within the program “Investissements d'Avenir” (ANR-11-IDEX-0007), operated 
by the French National Research Agency (ANR). This work was supported by a grant from the Simons
Foundation (651461, PPC).
It has been achieved thanks to the resources of PSMN from ENS de Lyon.

\newpage
\null
\newpage

\widetext

\section{Cascade of resonances to achieve internal wave turbulence: supplementary material} 

\centerline{G\'eraldine Davis$^1$, Timoth\'ee Jamin$^1$, Julie Deleuze$^1$, Sylvain Joubaud$^{1,2}$, Thierry Dauxois$^1$}
\centerline{\it 1. Univ Lyon, ENS de Lyon, Univ Claude Bernard, CNRS, Laboratoire de Physique, Lyon, France}
\centerline{\it 2. Institut Universitaire de France (IUF)}
\date{\today}

\bigskip

In this supplemental material, we present experimental details, movies, pictures, and additional
measurements of the internal wave attractor experiment. 
Notations are the same as in the above-mentioned paper.

\section*{1. Side view of the experimental set-up} 

A useful experimental technique to visualize internal wave attractors (IWA) is to mark many different isopycnals by a fluorescent dye, which allows us to follow their
\correc{displacement.}
This technique has been introduced  for the qualitative visualization of lee waves~\cite{Hopfingeretal1991}, and applied
later  to the qualitative visualization of stratified wakes and spin-up flows.
More recently, this technique has been shown to lead to quantitative  measurements using an algorithm to process
the experimental dye lines and determine their displacement to sub-pixel accuracy~\cite{VoisinErmanyukFlor2011}.

\begin{figure}[ht!]
\includegraphics[width=0.4\columnwidth,angle=270]{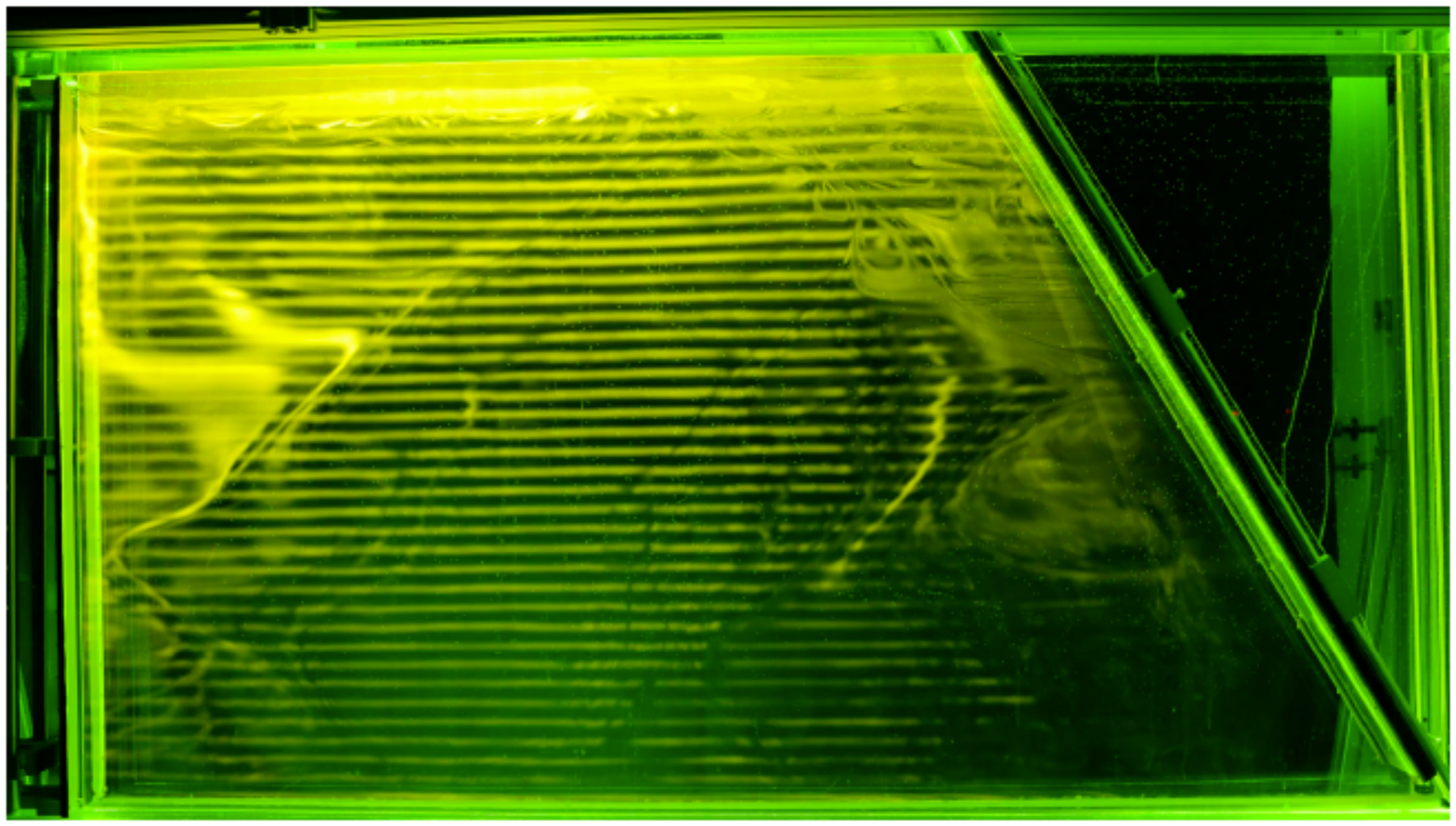}
\caption{View of the experimental tank: equidistant isopycnals have been emphasized with {rhodamine}. $N=0.77$ rad/s,  $\alpha={29.5}^\circ$ and $L=1600$~mm.}
\label{fig:Videos}
\end{figure}

Figure~\ref{fig:Videos} presents a side view of  the experimental tank. Approximately forty horizontal lines are visible from top to bottom. 
Once the wave generator on the left-hand side of the tank starts to move, internal gravity waves are  generated\correc{,} propagat\correc{e} in the tank,
reflect onto the slope\correc{,} and eventually converg\correc{e} towards an internal wave attractor.

The different movies (sped up by a factor $24$) correspond to different experimental parameters:\\
$\bullet$ Movie1.mp4: (1,1) internal wave attractor  obtained with $\omega_0/N = 0.53$  and $a=2$ mm.\\
$\bullet$ Movie2.mp4: (1,1) internal wave attractor  obtained with  $\omega_0/N= 0.53$  and $a=8$ mm.\\
 $\bullet$ Movie3.mp4: (2,1) internal wave attractor obtained  with  $\omega_0/N = 0.77$  and $a= 2$ mm.

\section*{2.  \gern{Energy} spectrum measurements during the kicked attractor experiment}

\begin{figure}[ht!]
\includegraphics[width=0.6\columnwidth]{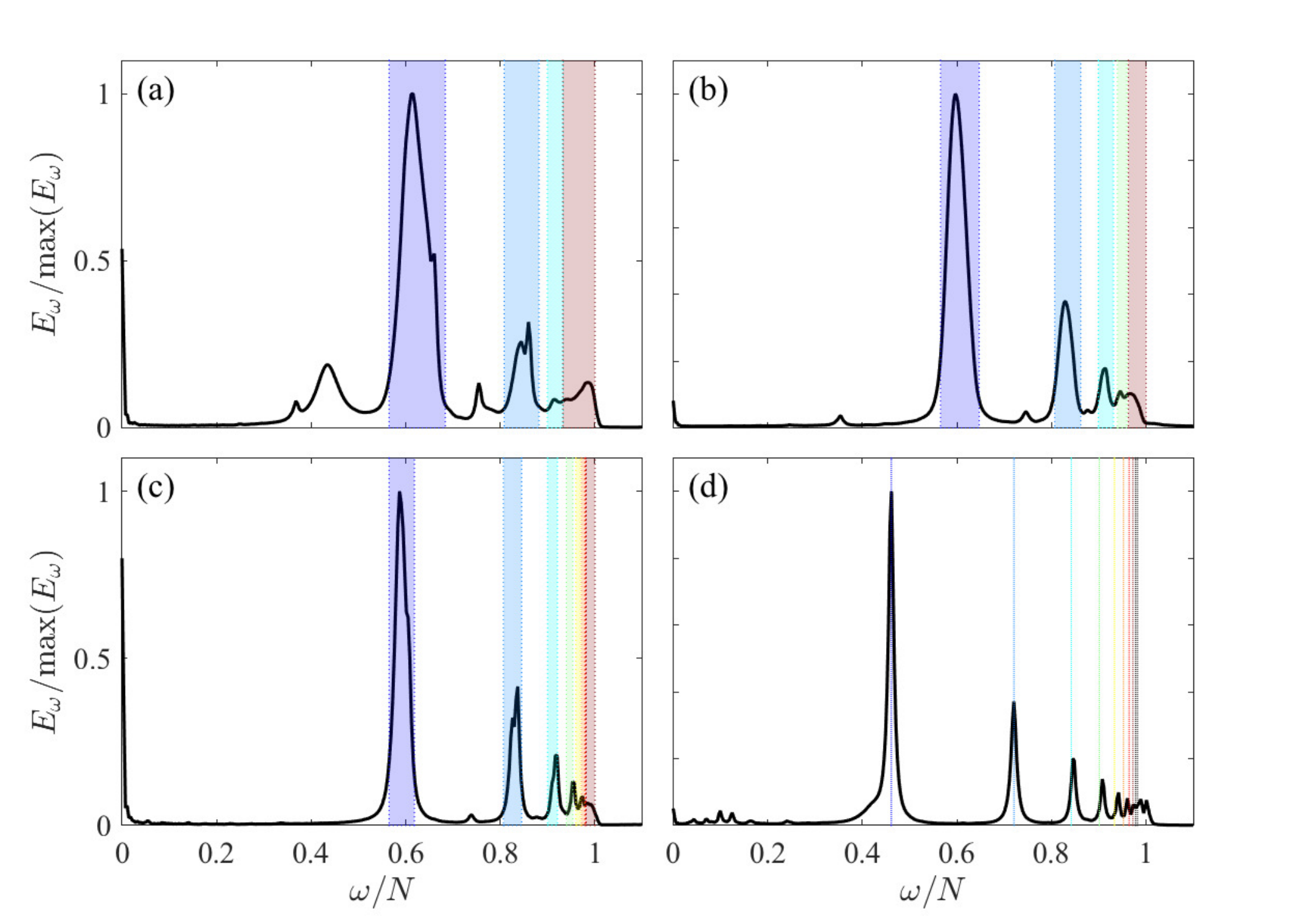}
\caption{Normalized \correc{energy spectral density} computed using the velocity components  $u(x,z,t)$ and $w(x,z,t)$  for $t \in [0,
1200]$ s, as in Fig.~\ref{fig:Lyapunov}\correc{(h).}
Panels (a), (b), (c) and (d) correspond to  $\alpha=21.4$, 15.5, 11.1 and 0$^\circ$,   \correc{with a maximum energy $6.3\cdot 10^{9}$~mm$^4$, $1.9\cdot 10^{9}$~mm$^4$,  $3.1\cdot 10^{9}$~mm$^4$ and $6.8\cdot 10^{9}$~mm$^4$ respectively}.  The length of the tank is $L=1328$~mm, except in (d) where $L=1750$~mm.
Vertical colored bands emphasize the theoretical frequency bands of internal wave attractor{s} in the stationary regime. 
Purple: attractors (1.1). Dark blue: attractors (2,1). Light blue: attractors (3.1). Green: attractors 
(4,1). Yellow: attractor (5.1). Orange: attractor (6.1). Red: point attractors. Panel (d) corresponds to the rectangular tank, 
where the theoretical resonance frequencies form a discrete set.}
\label{fig:Spectre}
\end{figure}

We compute the  \gern{energy} spectrum of the velocity field for different slope angles to highlight the effect of the tank geometry. 
These spectra are calculated over the entire duration of the experiment and are displayed in Fig.~\ref{fig:Spectre}. To limit the influence of noise, 
which is more important on the edges of the domain, the  \gern{integral} is performed 
over an area slightly smaller than the   experimental area. 
The colored regions correspond to the theoretical frequencies of the attractors. The color indicates the type of attractor as mentioned in the caption.
One can see that the \gern{energy} density is indeed larger on these frequency domains.

For the three different cases with a slope, 
panels (a-c)  show the existence of large bumps. 
On the other hand, for a rectangular tank, panel (d) shows that the  resonance frequencies form a discrete set. 
\correc{T}he non zero width of the peaks is due to dissipation effects.

\newpage

\newpage
\section*{3. Propagation of secondary waves generated via triadic resonant instability of an IWA}

It is known and now reasonably well documented that internal gravity waves are subject to  triadic resonant instability (TRI)~\cite{DauxoisARFM2018}.
Since an internal wave attractor is composed of four different branches that correspond to four internal gravity beams, it is not a surprise to 
see the destabilization of an IWA, 
especially \correc{of} its first branch after \correc{focusing} (in the upper right part of the tank) where the energy is greatest.

\begin{figure}[ht!]
\includegraphics[width=0.7\columnwidth]{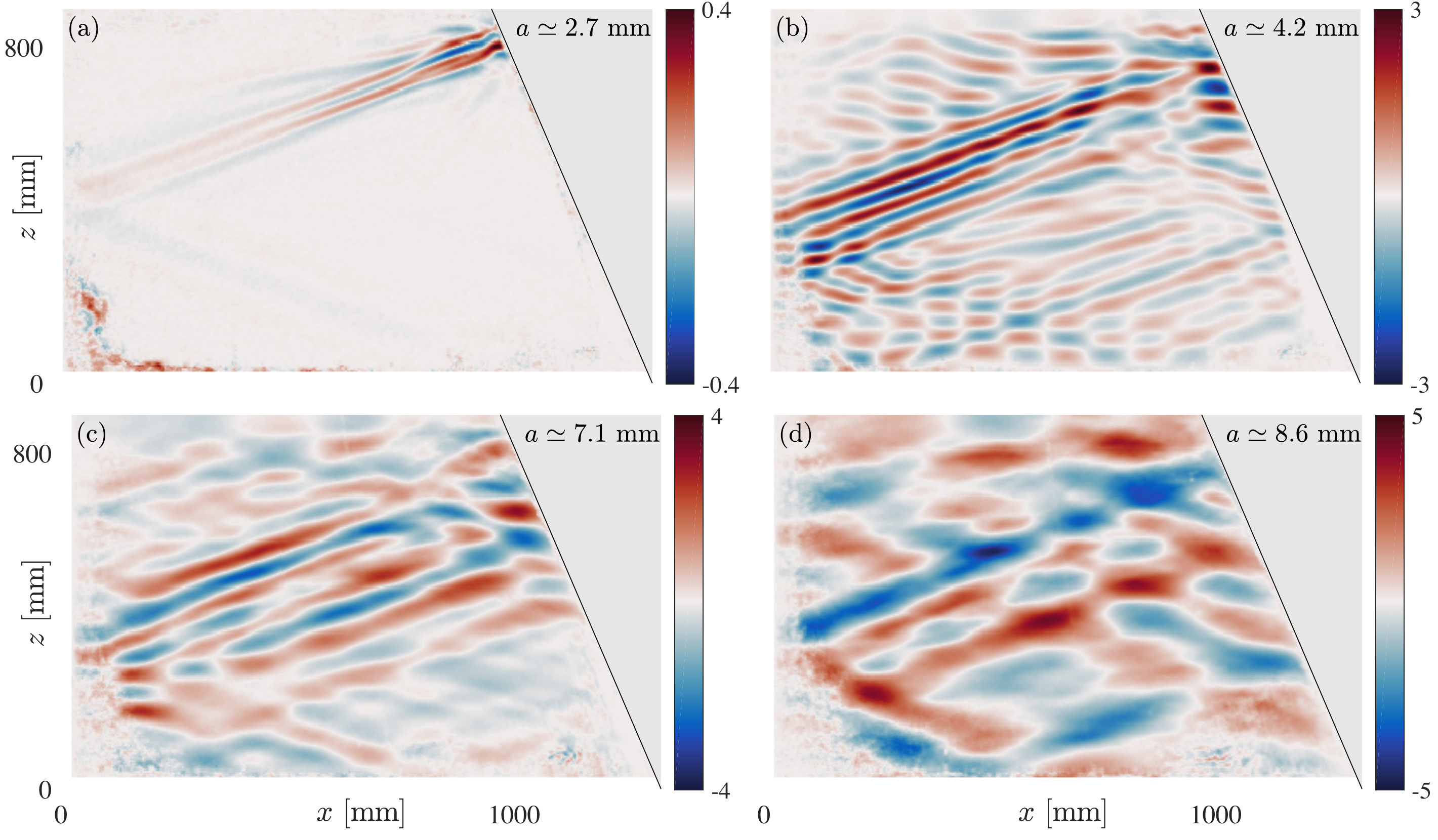}
\caption{Different instantaneous horizontal velocity fields (in mm/s), filtered \correc{around} the subharmonic frequency $\omega_+/N=0.37$.}
\label{fig:FilteredField}
\end{figure}

Figure~\ref{fig:FilteredField} presents the evolution \correc{as the forcing amplitude $a$ increases} of the horizontal velocity field filtered \gern{around} $\omega_+/N=0.37$ \correc{(\emph{i.e.}} around one of the two secondary-wave frequencies\correc{)}  when the tank is forced  at $\omega_0/N=0.62$. 
\correc{For the smallest forcing amplitude}, as shown by panel (a),  the field is intense only on the upper part of the tank, close to the first branch, and the pattern has a typical wavelength $\lambda \simeq 50$~mm. 
Then, after successive \correc{reflections}, the beam fills the entire tank (panel b). We can see that this phenomenon is accompanied by a gradual increase in wavelength, which reaches about $130$~mm in panels~(c) and~(d) .

\begin{figure}[ht!]
\includegraphics[width=0.8\columnwidth]{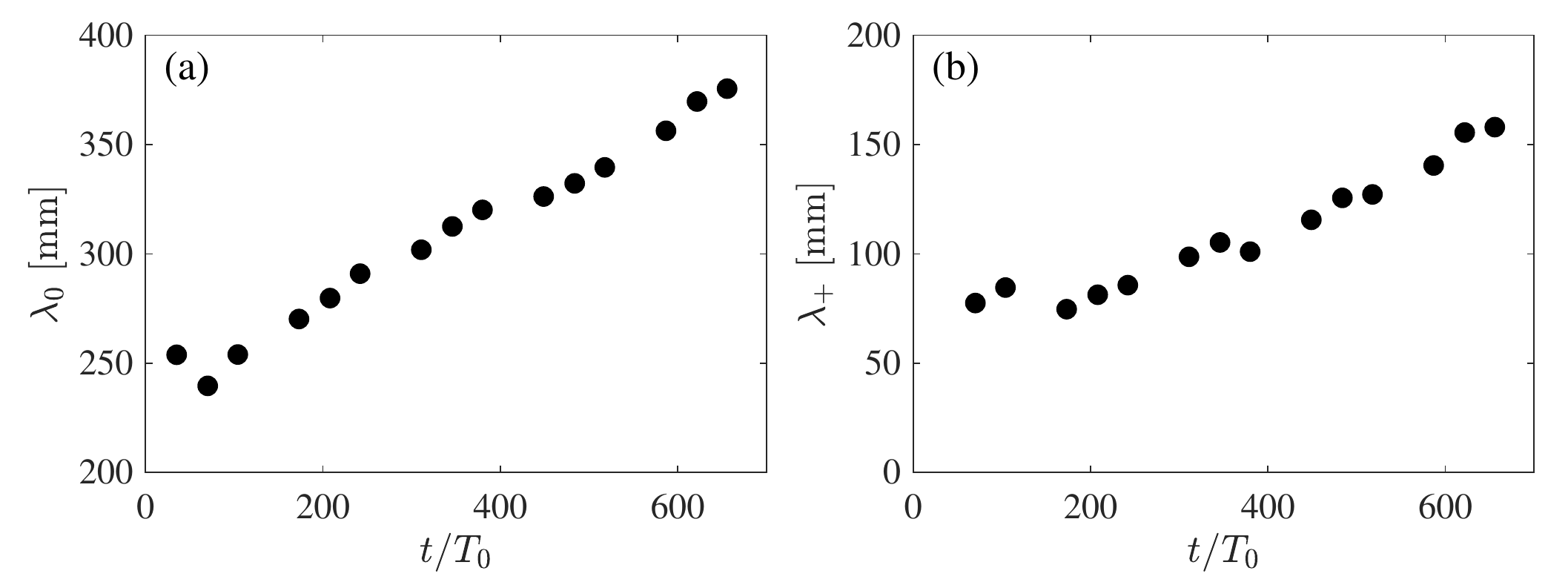}
\caption{Evolution of the dominant wavelength of the attractor for a forcing frequency  $\omega_0/N=0.62$~(a) and of one of the corresponding subharmonic wave  with a frequency close to  $\omega_+/N=0.37$~(b).}
\label{fig:Wavelength}
\end{figure}

Figure~\ref{fig:Wavelength} presents quantitative measurements to support this observation. Panel (a) shows the wavelength $\lambda_0$ of the second branch of the attractor over time.
This wavelength has been defined as the wavelength associated to the average wave vector~$\langle k_0 \rangle$, weighted by the  \correc{power spectral} density $E_{{\mathbf{k}}}$
\begin{equation}
\lambda_0 \equiv \frac{2\pi}{\langle k_0 \rangle} \qquad \mbox{with}\qquad
\langle k_0 \rangle \equiv 
\frac
{\displaystyle \int \mathrm{d}\mathbf{k} \, \vert \mathbf{k} \vert E_{{\mathbf{k}}}(\mathbf{k}, \omega_0)}
{\displaystyle \int \mathrm{d}\mathbf{k}   \,                               E_{{\mathbf{k}}}(\mathbf{k}, \omega_0)},
\label{eq:lambda0}
\end{equation}
where $ E_{\mathbf{k}}(\mathbf{k}, \omega_0)$ is computed from the velocity field filtered at the forcing frequency $\omega_0$, and the integrals over $\mathbf{k}$ are computed only for $k_x<0,k_z>0$ (which corresponds to the second branch of the attractor).
We see a $50\%$ increase of this wavelength, between the beginning and the end of the experiment, since  $\lambda_0$ rises from approximately 250 to 380~mm. 
In the same way, we can compute the characteristic wavelength $\lambda_+$ of the secondary wave presented in Fig.~\ref{fig:FilteredField} using Eq.~(\ref{eq:lambda0}), where the velocity field is filtered   \gern{around} $\omega_+/N=0.37$. 
This increase of  the wavelength of the primary wave is clearly correlated with an increase of the wavelength of the considered secondary wave, as shown in 
 panel~\correc{\ref{fig:Wavelength}}(b): the latter doubles between the beginning and the end of the experiment 
 As the group velocity of a wave is proportional  to its wavelength, it also doubles from 8 to $16$~mm/s. 
 The distance travelled during the characteristic dissipation time  (of the order of $150$~s) thus 
increases  from 1 to $2$~m.  The beam can interfere with itself and possibly present a quasi-stationary structure.

We just saw that in larger forcing cases \correc{(Fig.~\ref{fig:Wavelength})}, the attractor is thicker
and the secondary wavelength is larger: consequently, the secondary waves fill available space and sometimes look like standing waves. 
A weakly focused attractor would moreover favor this effect since it already has a longer wavelength in the linear regime. 
The quasi-stationarity of the secondary waves is thus related to the wavelength of the attractor. This interpretation is different from the one suggested in~\cite{Brouzetetal2017}, which described it as a consequence of the quasi-stationarity of the attractor.

\begin{figure}[ht!]
\includegraphics[width=0.4\columnwidth]{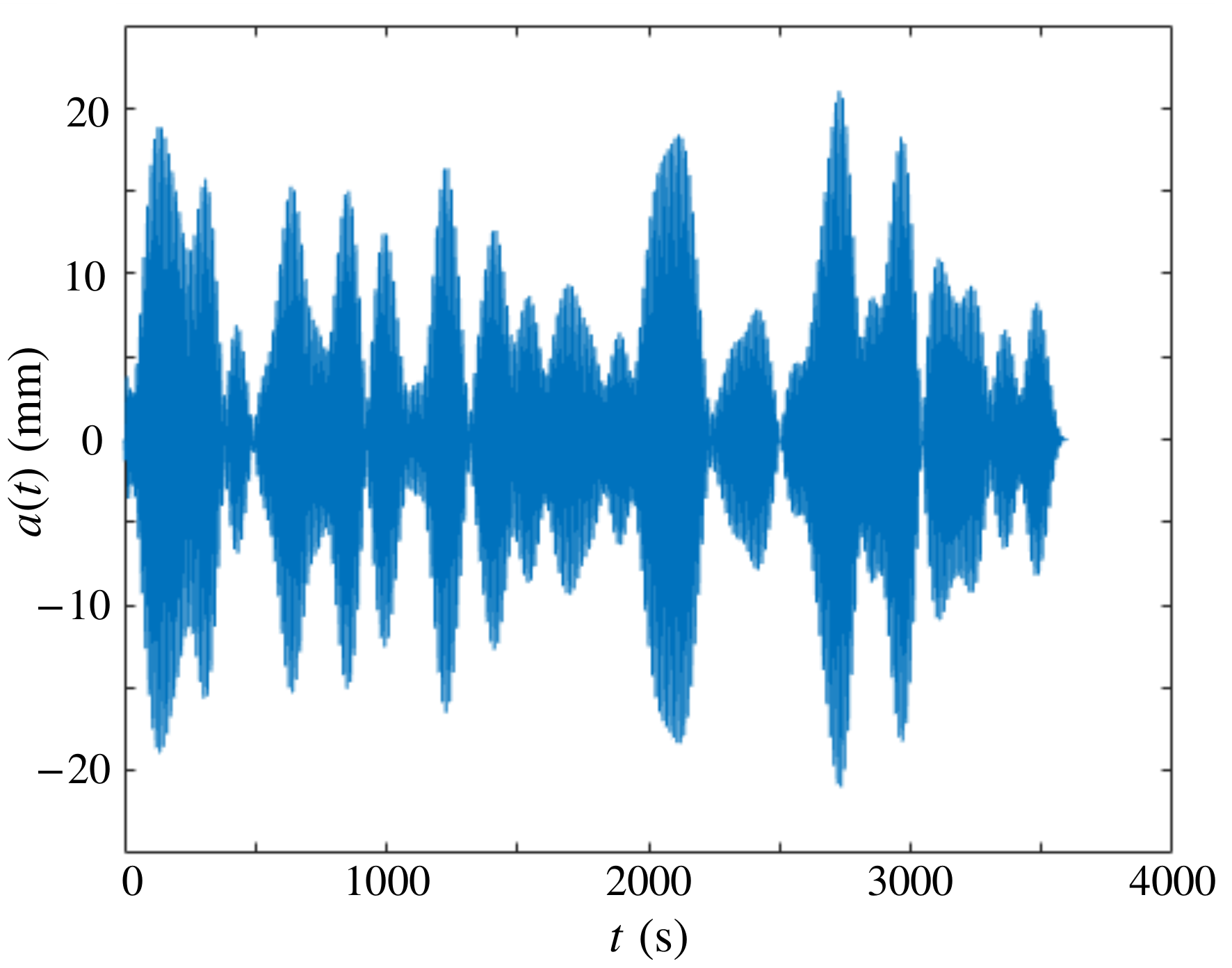}
\caption{\correc{Amplitude signal used for the frequency modulated forcing.}}
\label{fig:typ_amplitude_freq_mod}
\end{figure}

\correc{\section*{4. Experimental parameters}
Table~\ref{table:exp_parameters} gives the experimental parameters used for the different stationary spectrum shown in Fig.~\ref{fig:SpatialSpectrum}(b). For the frequency modulated forcing, the amplitude signal is given in Fig.~\ref{fig:typ_amplitude_freq_mod} and is characterized by its standard deviation.}
\begin{table*}[h]
\begin{center}
\begin{tabular}{c||c|c|c|c}
\hline \hline
~Forcing~ & ~Wave period $T_0$ (s)~ & ~Amplitude $a$~(mm)~ & ~angle $\alpha$ ($^{\circ}$)~ & ~$L$~(mm)~ \\
\hline \hline
weakly focused & 14.5 & growth from 8.3~mm to 9.4~mm &21.4 & 1328 \\
strongly focused    & 13.6  & 14& 25.3 & 1455\\
frequency modulated  & [12.9 - 14.9]    & ${\langle a^2\rangle}^{1/2} = 7$~mm & 15.7 & 1340\\
\hline \hline 

\end{tabular}
\caption{\correc{Experimental parameters for Fig.~\ref{fig:SpatialSpectrum}(b)}\label{table:exp_parameters}}
\end{center}
\end{table*}

\correc{\section*{5. Nonlinear time scale}
The wave turbulence theory is developed within a framework of weak turbulence, where nonlinear interactions induce wave modulation with a typical time scale \---~called the nonlinear time scale and denoted $\tau_{NL}$~\--- that has to be long enough compared to the wave period $T=2\pi/\omega$. 
}

\correc{
\timn{Let us} consider the \tim{simple case of a} weakly resonant interaction between three monochromatic plane waves $i$, $j$ and $\ell$. Assuming that the wave amplitudes vary slowly with respect to the periods of the waves, straightforward calculations~\cite{BourgetJFM2013,DauxoisARFM2018} 
lead, at first order, to \thie{three} amplitude equations for the stream function\thie{s} \timn{similar to} 
\begin{equation}
\quad \frac{\textrm{d}\psi_{\thie{\ell}}}{\textrm{d}t} = I_{\thie{\ell}} \psi_{\thie{i}}\psi_{\thie{j}} - \frac{\nu}{2} k_{\thie{\ell}}^2 \psi_{\thie{\ell}}\,,
\label{Eq:evol_amp}
\end{equation}
where $I_{\thie{\ell}}$ is a function of the characteristics of the \thie{triad} \thie{$(i,j,\ell)$} in interaction. 
\tim{The viscous time scale $\tau_{\nu}\approx 1/(\nu k_{\thie{\ell}}^2)\in[100;2500]$\,s can be immediately deduced from Eq.~(\ref{Eq:evol_amp}).}}

\correc{On the one hand, \tim{these equations lead} to triadic resonance \emph{instability} when the amplitude of one wave is \tim{large enough to generate} 
 two other ones \timn{(with a viscous threshold, see~\cite{BourgetJFM2013,DauxoisARFM2018})}. 
On the other hand, \thie{without any threshold,} two waves 
can lead to the growth of a third wave corresponding to triadic \emph{interaction} \timn{(which comes directly from Eq.~(\ref{Eq:evol_amp}) by considering $\psi_i$ and $\psi_j$ constant).}
} 
\timn{These nonlinear mechanisms come with timescales that we estimate experimentally.}

\begin{figure}[ht!]
\includegraphics[width=0.8\columnwidth]{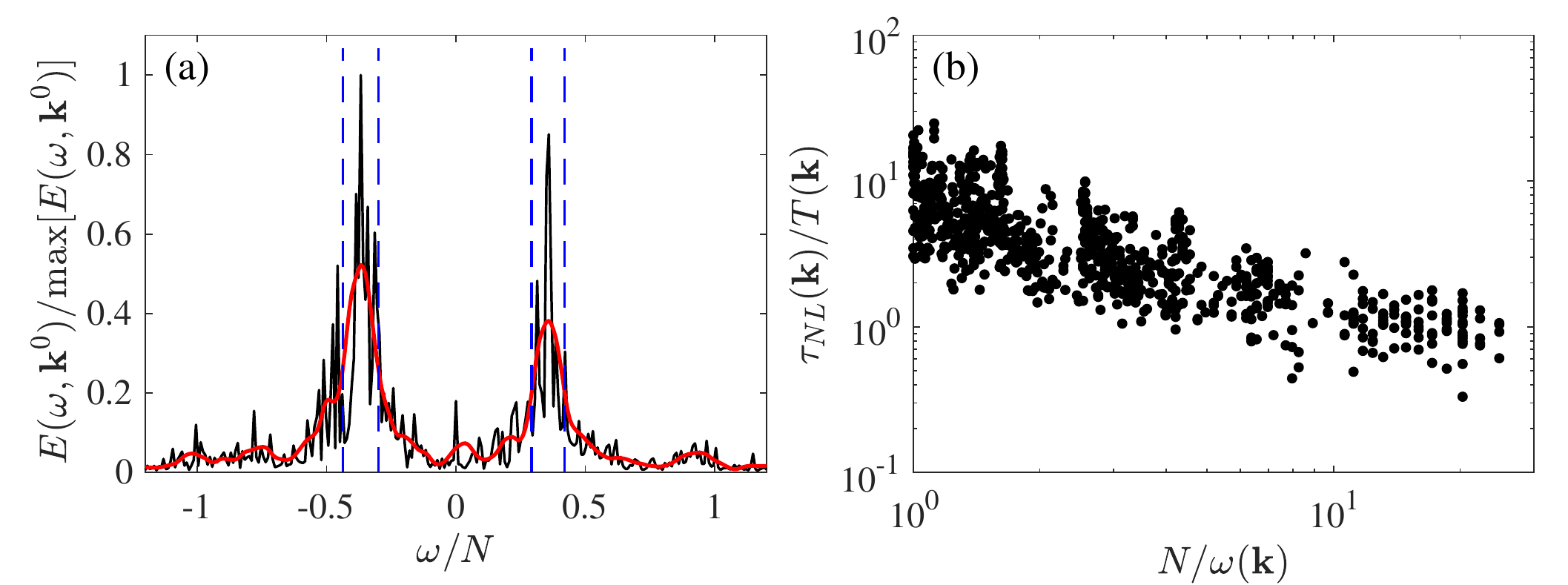}
\caption{\correc{
(a) \thie{Temporal Fourier } spectrum \timn{taken from} the power spectrum  $E(\omega,\mathbf{k})$ for \thie{the} given wavevector $(k_x^0, k_z^0)=(-0.15, 0.63)$~rad/m (thin black lines). The thick red curve is a smoothed curve. The dashed vertical lines indicate the width at mid-depth of each peak. 
(b) \tim{Experimental normalized} nonlinear time $\correc{\tau_{NL}}$ given by the width at mid-depth \tim{of the peaks from (\thie{a})-like spectra} 
as a function of the \tim{normalized wave frequency}.
}}
\label{fig:TNL_spectrum}
\end{figure}

\begin{figure}[ht!]
\includegraphics[width=0.8\columnwidth]{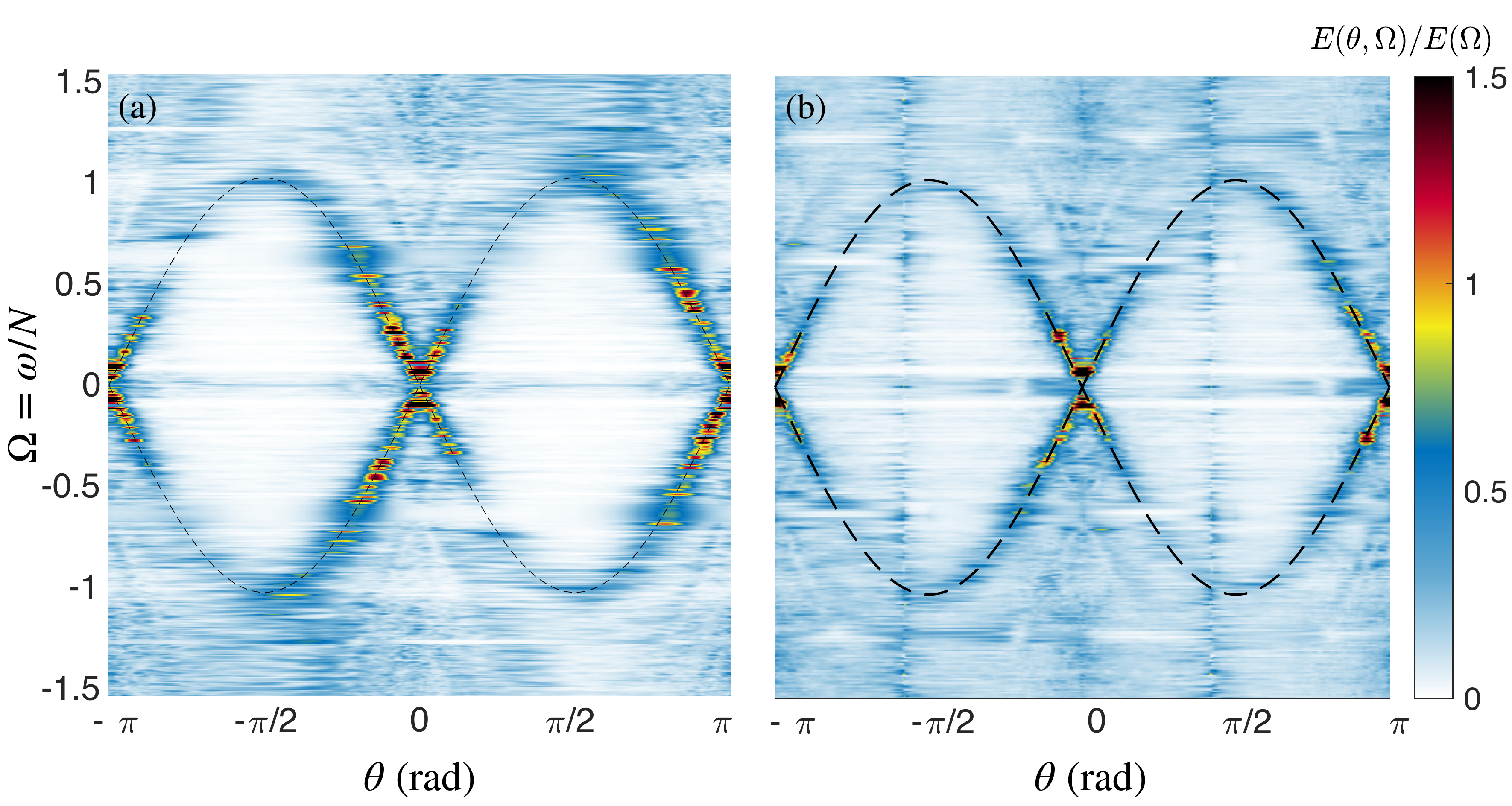}
\caption{\correc{Normalized power spectrum $E(\theta,\omega)/(\int_0^{2\pi} E(\theta,\omega)\textrm{d}\theta)$ for (a) the weakly focused case and (b) the frequency modulated case. Dashed lines correspond to the dispersion relation of internal waves $\omega/N=\sin\theta$.}}
\label{fig:Disp_relation}
\end{figure}

\correc{\tim{As recently shown by Yarom, Salhov and Sharon~\cite{Yarom2017}, the nonlinear time scale is related to the width of peaks in temporal spectra, as the one represented in Fig.~\ref{fig:TNL_spectrum}(\thie{a}), taken from the power spectrum at a given wavevector $\mathbf{k^0}$. The nonlinear time associated to this wavevector is then defined by $\tau_{NL}= 2\pi/\Delta\omega$, where $\Delta\omega$ is the mean width at mid-depth of the two peaks around $\omega=\pm N \sin \theta_{\mathbf{k^0}}$. The ratio between the nonlinear and the linear times is then represented in Fig.~\ref{fig:TNL_spectrum}(\thie{b}) for various wavevectors \tim{in the inertial range ($k\in[20; 100]$\,rad/m, see Fig.~\ref{fig:SpatialSpectrum})}. 
A large majority of the values for $\tau_{NL}/T$ \thie{are} lying between 1 and~10.} This order of magnitude does not correspond to the theoretical framework of \correc{\it weak} wave turbulence \correc{($\tau_{NL}/T \gg 1$)}. \tim{Nevertheless, as shown by Fig.~\ref{fig:Disp_relation} displaying the normalized power spectra as a function of $\theta$ and $\omega$,} the energy is localized around the dispersion relation of internal waves \tim{(dashed lines)} for all cases. \tim{This confirms we actually observe multiple internal wave interactions leading to wave turbulence. }}

\end{document}